\documentclass{PoS}

\newcommand{\I}{{\scriptscriptstyle I}}

\def\dags{{\protect\mbox{\tiny \dag}}}
\usepackage{float}
\usepackage{bm}
\usepackage{amsmath}
\title{Constraints on generalized non-standard \bm{$tbW$} couplings}

\ShortTitle{Constraints on generalized non-standard $tbW$ couplings}

\author{Zenr\=o Hioki\\
        Institute of Theoretical Physics,\\ University of Tokushima, Tokushima\\
        E-mail: \email{hioki@tokushima-u.ac.jp}}

\author{\speaker{Kazumasa Ohkuma}\\
        Department of Information and Computer Engineering,\\ Okayama University of Science, Okayama\\
        E-mail: \email{ohkuma@ice.ous.ac.jp}}

\author{Akira Uejima\\
        Department of Information and Computer Engineering,\\ Okayama University of Science, Okayama\\
        E-mail: \email{uejima@ice.ous.ac.jp}}
\abstract{
General non-standard $tbW$ couplings are studied as model independently as possible based on
the effective Lagrangian consisting of the dimension-6 operators, which is an extension of the standard-model
Lagrangian. The $tbW$-interaction Lagrangian in this framework includes four kinds of couplings, which could be
complex. Constraints on those non-standard $tbW$ couplings are obtained by comparing the experimental data
related to the $t\to b W$ process with the corresponding theoretical formulas derived from the effective Lagrangian.
The constraints on some sets of the non-standard couplings are shown not to be so strong because those couplings
balance out each other as we treat all the non-standard couplings as complex numbers at the same time.
}

\FullConference{38th International Conference on High Energy Physics\\
		3-10 August 2016\\
		Chicago, USA}

\begin{document}
The top quark, discovered more than 20 years ago, is regarded as one of the most attractive particles for
new-physics searches even now because of its huge mass.
Therefore, top-quark physics is one of the main topics at the High-Luminosity Large
Hadron Collider (HL-LHC) and the International Linear Collider (ILC) as well as the current Large Hadron Collider (LHC).
In particular, various measurements of the top quark will be performed more precisely at the HL-LHC and the ILC as
new-physics searches.
At the experiments the purpose of which is precision measurements, deviations from 
the standard-model predictions are probed as signals of new physics. However, it is not easy to specify the model
for the new physics from the beginning since such deviations could be induced via quantum effects of new particles
in many other models as well.
Thus, model-independent analyses would be an important key to building new-physics models beyond the standard model 
as a bottom-up approach.
In this proceedings, we report the constraints on the most general non-standard $tbW$ couplings
in the effective Lagrangian based on our two recent studies~\cite{Hioki:2015env,Hioki:2016xtc}.

In order to construct the effective Lagrangian, we assume the masses of non-standard particles are much heavier
than the cut-off scale ${\mit \Lambda}$. In this case, the effective Lagrangian is written as
\begin{equation}
{\cal L}_{\rm eff} ={\cal L}_{\rm SM} +\frac{1}{{\mit \Lambda}^2}\sum_i\left[ C_i {O}_i +  C_i^* {O}_i^\dagger\right],
\end{equation}
where ${\cal L_{\rm SM}}$ is the standard-model Lagrangian, ${O}_i$ are the dimension-6 operators and
their unknown coefficients $C_i$, combined with
${\mit \Lambda}^{-2}$, produce the non-standard coupling constants.
The $t\to b W$ process we focus on could be affected by the following
operators~\cite{AguilarSaavedra:2008zc, Grzadkowski:2010es};
\begin{eqnarray}
&&{O}^{(3,33)}_{\phi q}
=i\sum_{\I}\,[\,\phi^{\dags}(x)\tau^{\I} D_\mu \phi(x)\,] 
[\,\bar{q}_{L3}(x)\gamma^\mu \tau^{\I} q_{L3}(x)\,]  \\
&&{O}^{33}_{\phi \phi}
=i[\,\tilde{\phi}^{\dags}(x) D_\mu \phi(x)\,]
[\,\bar{u}_{R3}(x)\gamma^\mu d_{R3}(x)\,]   \\
&&{O}^{33}_{uW}
=\sum_{\I} \,\bar{q}_{L3}(x) \sigma^{\mu\nu} \tau^{\I} u_{R3}(x)
\tilde{\phi}(x) W^{\I}_{\mu\nu}(x)~~~~~~   \\
&&{O}^{33}_{dW}
=\sum_{\I} \:\bar{q}_{L3}(x) \sigma^{\mu\nu} \tau^{\I} d_{R3}(x)
\phi(x) W^{\I}_{\mu\nu}(x),
\end{eqnarray}
where the notations obey basically those in Ref.\cite{AguilarSaavedra:2008zc}.
Using these operators, we derive the effective $tbW$ Lagrangian as
\begin{eqnarray}
&&{\cal L}_{tbW} =-{g\over\sqrt{2}}\:\Bigl[\,
\bar{\psi}_b(x)\gamma^{\mu}(f_1^L P_L +f_1^R P_R)\psi_t(x) W_\mu^-(x) \nonumber \\
&&\phantom{{\cal L}_{tbW} =-{g\over\sqrt{2}}}\ \ \
+\,\bar{\psi}_b(x){{\sigma^{\mu\nu}}\over M_W}
(f_2^L P_L +f_2^R P_R)\psi_t(x) \partial_\mu W_\nu^-(x)\,\Bigr],
\label{LagW}
\end{eqnarray}
where $g$ is the $SU(2)$ coupling constant, $P_{L/R}\equiv(1\mp\gamma_5)/2$,
\[
\begin{array}{ll}
f_1^L \equiv V_{tb}+C_{\phi q}^{(3,33)*} \displaystyle{\frac{v^2}{{\mit\Lambda}^2}},
&\ \ \ \
f_1^R \equiv C_{\phi \phi}^{33*} \displaystyle{\frac{v^2}{2{\mit\Lambda}^2}}, \\
f_2^L \equiv -\sqrt{2}C_{dW}^{33*} \displaystyle{\frac{v^2}{{\mit\Lambda}^2}},
&\ \ \ \
f_2^R \equiv -\sqrt{2}C_{uW}^{33} \displaystyle{\frac{v^2}{{\mit\Lambda}^2}}, \\
\end{array}
\]
with $v$ being the Higgs vacuum expectation value and $V_{tb}$ being the $(tb)$ element of
Kobayashi-Maskawa matrix. We then decompose $f_1^L$ into the standard  and non-standard model parts: 
$f_1^L =f_{\rm SM}^1 +\delta f_1^L$, and set as $f_{\rm SM}^1=V_{tb}=1$ and
 $\delta f_1^L\equiv C_{\phi q}^{(3,33)*} \displaystyle{{v^2}/{{\mit\Lambda}^2}}$
hereafter.

Although studies of the $tbW$ couplings using the effective Lagrangian
have already been performed in order to probe possible new interactions
\footnote{The preceding works are listed in~\cite{Hioki:2015env, Hioki:2016xtc}, and here we
add Refs.~\cite{Cirigliano:2016njn, Cirigliano:2016nyn, Boos:2016zmp} to the lists.},\
it is assumed there that the non-standard couplings are real numbers, or partially complex numbers,
and/or only some couplings have been treated as free parameters at once fixing the others. 
In addition, it has not been unusual to adopt the linear approximation in those parameters,
i.e., to neglect their quadratic (and higher-power) terms.
However, in this analysis, we give the constraints on non-standard $tbW$ couplings treating
all those couplings as complex numbers  (i.e., an eight-parameter analysis is carried out)
without the above-mentioned assumption and approximation.

In our numerical analysis, the following experimental information is used as our input data:\\
\hspace*{1cm}$\bullet$ the total decay width of the top quark~\cite{Khachatryan:2014nda}
\begin{equation}\label{eq:total_w}\vspace*{-0.2cm}
{\mit\Gamma}^t = 1.36\pm 0.02({\rm stat.})^{+0.14}_{-0.11}({\rm syst.}) ~~{\rm GeV}.
\footnote{
    In fact, it is not easy to handle an asymmetric error like this in the error propagation,
    we use ${\mit\Gamma}^t = 1.36\pm 0.02({\rm stat.})\pm 0.14({\rm syst.}) ~{\rm GeV}$, the one
    symmetrized by adopting the larger (i.e., $+0.14$) in this systematic error.}
\end{equation}
\hspace*{1cm}$\bullet$ the partial decay widths derived from experimental data of $W$-boson helicity
fractions~\cite{Khachatryan:2016fky}
\begin{equation}\label{eq:gamma_eff}\vspace*{-0.2cm}
\begin{split}
 &{\mit\Gamma}_L^{t*}=0.439\pm 0.051~{\rm GeV},\\
 &{\mit\Gamma}_0^{t*}=0.926\pm 0.103 ~{\rm GeV},\\
 &{\mit\Gamma}_R^{t*}=-0.005\pm0.020~{\rm GeV}.
\end{split}
\end{equation}
Varying all the non-standard couplings at the same time, we have compared the above experimental
data with the corresponding theoretical formulas derived from the effective Lagrangian, and explored
allowed areas for each parameter.
The resultant constraints on the eight parameters are shown in Table \ref{tab:8para}.
In addition, since it might seem strange that the contribution from the standard-model coupling 
$f_{\rm SM}^1$ is diminished by its extended coupling $\delta f_L^1$ as can be seen in Table\ref{tab:8para},
the constraints on the other couplings in the case of 
Re($\delta f_1^L$) = 0 and Re($\delta f_1^L$) = Im($\delta f_1^L$) = 0 are also derived and 
shown in Tables \ref{tab:7para} and \ref{tab:6para}.

\begin{table}[H]
\centering
\caption{Allowed maximum and minimum values of the non-standard-top-decay couplings in the case
that all the couplings are dealt with as free parameters.
}
\label{tab:8para}
\vspace*{0.4cm}
\begin{tabular}{c|cc|cc|cc|cc}
& \multicolumn{2}{c|}{$\delta\! f_1^L$}& \multicolumn{2}{c|}{$f_1^R$}
& \multicolumn{2}{c|}{$f_2^L$}& \multicolumn{2}{c}{$f_2^R$}
\\ \cline{2-9} 
& Re($\delta\! f_1^L$)
&\hspace*{-0.4cm} Im($\delta\! f_1^L$) & Re($f_1^R$)
&\hspace*{-0.4cm} Im($f_1^R$) & Re($f_2^L$)
&\hspace*{-0.4cm} Im($f_2^L$) & Re($f_2^R$)
&\hspace*{-0.4cm} Im($f_2^R$)
\\ \cline{1-9}
Min. & $-2.58$    &\hspace*{-0.4cm} $-1.58$           & $-1.36$           
&\hspace*{-0.4cm} $-1.36$           & $-0.68$           
&\hspace*{-0.4cm} $-0.68$           & $-1.20$            
&\hspace*{-0.4cm} $-1.20$           \\ 
Max. & $\phantom{-}0.58$ &\hspace*{-0.4cm} $\phantom{-}1.58$ & $\phantom{-}1.36$ 
&\hspace*{-0.4cm} $\phantom{-}1.36$ & $\phantom{-}0.68$ 
&\hspace*{-0.4cm} $\phantom{-}0.68$  & $\phantom{-}1.20$  
&\hspace*{-0.4cm} $\phantom{-}1.20$ \\
\end{tabular}
\end{table}


\begin{table}[H]
\centering
\caption{Allowed maximum and minimum values of non-standard-top-decay couplings in the case
that all the couplings are dealt with as free parameters except for Re$(\delta\! f_1^L)$ being
set to be zero.
}
\label{tab:7para}
\vspace*{0.4cm}
\begin{tabular}{c|c|cc|cc|cc}
& {$\delta\! f_1^L$}& \multicolumn{2}{c|}{$f_1^R$}& \multicolumn{2}{c|}{$f_2^L$}
& \multicolumn{2}{c}{$f_2^R$}             \\ \cline{2-8}
&  Im($\delta\! f_1^L$) & Re($f_1^R$) & Im($f_1^R$) & Re($f_2^L$) &Im($f_2^L$) 
& Re($f_2^R$) & Im($f_2^R$) \\  \cline{1-8}
Min. 	&  $-1.23$  & $-1.14$ 
	& $-1.12$ & $-0.55$ 
        & $-0.57$ & $-0.96$ 
        & $-1.00$ \\ 
Max. 
& $\phantom{-}1.23$  
& $\phantom{-}1.10$ & $\phantom{-}1.12$
& $\phantom{-}0.59$  & $\phantom{-}0.57$
& $\phantom{-}0.00$ & $\phantom{-}1.00$\\
\end{tabular}
\end{table}


\begin{table}[H]
\centering
\caption{Allowed maximum and minimum values of non-standard-top-decay couplings in the case
that all the couplings are dealt with as free parameters except for Re$(\delta\! f_1^L)$ and
Im$(\delta\! f_1^L)$ both being set to be zero.
}
\label{tab:6para}
\vspace*{0.4cm}
\begin{tabular}{c|cc|cc|cc}
& \multicolumn{2}{c|}{$f_1^R$}& \multicolumn{2}{c|}{$f_2^L$}
& \multicolumn{2}{c}{$f_2^R$} \\ \cline{2-7} 
&  Re($f_1^R$) & Im($f_1^R$) & Re($f_2^L$) &Im($f_2^L$) & Re($f_2^R$) & Im($f_2^R$) \\ \cline{1-7}
Min. 	&  $-1.14 $ & $-1.12$ 
 	& $-0.55$ & $-0.57$ 
        & $-0.96$ & $-0.49$  \\ 
Max. 	&  $\phantom{-}1.10$   & $\phantom{-}1.12$ 
	&  $\phantom{-}0.59$   & $\phantom{-}0.57$
     	&  $\phantom{-}0.00$   & $\phantom{-}0.49$\\
\end{tabular}
\end{table}

In summary, the maximum and minimum values of the non-standard $tbW$ couplings allowed by the present experimental
data of the top-quark total and partial decay widths were  derived by varying all the couplings independently
at the same time. We found that
\begin{itemize}
 \item In the case that all the coupling constants are treated as complex numbers,
 the allowed regions of those couplings are not that small yet because cancellations could
happen among the contributions originated from those couplings.
 \item If we assume that $f_1^L$ does not include any non-standard contribution, the resultant
constraints on the other non-standard couplings, especially $f_2^R$, become a bit stronger,
although their allowed ranges are not such tiny that we can drop their quadratic terms
safely.
\end{itemize}

\end{document}